\documentclass[11pt]{ucscthesisbs}
\usepackage[utf8]{inputenc}
\usepackage[labelfont=bf]{caption}
\usepackage{subfig}
\usepackage{graphicx}
\usepackage{mathtools, siunitx}
\usepackage{indentfirst}
\usepackage{textcomp}
\usepackage{hyphenat}

\begin{document}
\title{Analyzing Variation in Phase Delays Across Phase Plates
With a Quadrature Polarization Interferometer 
}
\author{Cesar Laguna}
\degreeyear{2021}
\degreemonth{27 August}
\degree{BACHELOR OF SCIENCE}
\field{ASTROPHYSICS}%

\chair{Bruce Schumm}      
\technicaladvisor{Lab of Adaptive Optics Director Philip Hinz}
\campus{Santa Cruz}
\maketitle
\copyrightpage

\begin{frontmatter}

\begin{abstract}
In order for telescopes to obtain good and precise images they need to see through atmospheric turbulence. To accomplish this and compensate for atmospheric turbulence we use Adaptive Optics technologies. In this thesis we analyze the variations in phase delays across phase plates which simulate atmospheric turbulence in order to characterize them and determine how well these phase plates reproduce the phase delay variation of the atmosphere. This experiment was conducted using the Quadrature Polarization Interferometer (QPI) testbed in the Lab of Adaptive Optics (LAO) at the University of California Santa Cruz (UCSC). Using the QPI lab setup allowed us to be able to develop and refine a final algorithm for determining the phase delay of the phase plates. The characterization of the phase plates was accomplished by using the interference patterns between a test and reference path and measuring their pathlength variations. This was achieved by calculating the intensity of the two paths of the Helium-Neon (HeNe) laser, modifying those values in order to obtain phase values between the beams, and from the phase values we are able to determine the path length variations, or phase delays, of the phase plates. This allows us to determine Fried’s parameter, $r_o$, by analyzing any given separation on the phase plate and determining the path length variations between those separations. 
\end{abstract}

\tableofcontents{}
\listoffigures{}
\listoftables{}

\begin{dedication}
\null\vfil
{\large
\begin{center}
To my parents,\\\vspace{12pt}
Cesar and Sarai Laguna
\end{center}}
\vfil\null
\end{dedication}

\begin{acknowledgements}
Thank you to my thesis advisor, family and friends for all their support and help throughout not only this thesis but also my academic career.
\end{acknowledgements}

\end{frontmatter}

\chapter{Introduction}
\section{Background: Adaptive Optics in Astronomy}

Atmospheric turbulence or wavefront distortion is produced by light passing through Earth’s atmosphere which has turbulent air flow thus the name atmospheric turbulence. This distortion is caused by a varying refractive index which in turn is brought about by temperature gradient, the direction and rate at which temperature changes in a given area. The difference in temperature creates different delays in the light passing through it which results in atmospheric turbulence. This atmospheric turbulence, “seeing”, affects the focus or sharpness of an image; the “twinkling” of stars is a byproduct of these distortions. Adaptive Optics (AO) is a technology that is used to compensate for atmospheric turbulence. AO removes wavefront distortions by using “counter wavefront distortions” in the form of a deformable mirror which makes corrections to the wavefront distortions in real time and is also able to “extends diffraction\hyp{limited} imaging to much fainter and complex objects” (Beckers,14). 

The resolution of optical instruments is primarily limited by the diffraction limit, the resolution power of an optical imaging system; even a “perfect” lens will be limited by its diffraction limit. Assuming no disturbances (turbulence, noise, etc) or imperfections in the mirror the diffraction limit is given by: 
\begin{equation}
    \theta = 1.22 * \frac{\lambda}{D}
\end{equation}
 where $\theta$ is the resolvable angle, $\lambda$ is the wavelength of the light, and D is the diameter of the lens. However, due to the effect that atmospheric turbulence produces, D is now replaced with $r_o$ or Fried’s parameter. Fried’s parameter, $r_o$, is the measurement of the strength of optical transmission through turbulent media due to the differences in temperature gradient in the atmosphere; inhomogeneities in the atmospheres refractive index caused by temperature gradients. More precisely it is defined as the size on the wavefront, usually the diameter of a circular aperture/area, where the root-mean-square variation (phase) is 1 radian. By replacing D with $r_o$, the angular resolution is now limited to approximately: 
\begin{equation}
    \theta = \frac{\lambda}{r_o}
\end{equation}

Since the diameter for most telescopes today is D$>>r_o$, AO must be implemented in order to obtain image resolution results that approach the diffraction limit. AO allows us to modify telescopes and fine tune them until we are able to to reach the diffraction limit of a lens and thus be able to obtain sharp images of the universe without the worry of turbulence. 

\section{Testbed}
To begin our testing we want to replicate and simulate atmospheric turbulence, or $r_o$, in the lab. By simulating an atmosphere in the lab we are in a position to test and refine AO techniques with more ease. The precision of the simulation relies on how large or small our replicated $r_o$ is. In the Earth’s atmosphere the $r_o$ is relatively small because of the tiny variations in temperature in small areas.  When comparing the small replicated/simulated $r_o$ it is not the same “small” $r_o$ that is seen in the atmosphere. In the atmosphere $r_o$ is typically 100 - 200mm however this is with our various large telescopes, such as the Keck telescope which is 10 meters in diameter. When simulating this in the lab environment we are constructing and using an optical setup which is quite smaller. This means that when we simulate atmospheric turbulence it will be far more convenient if the “telescope”, or beam, is only 2mm in size which in turn means that the $r_o$ we want to simulate should range between .5 - 1.5mm in size, keeping the ratio consistent $r_o$ should be 1mm in size. 

In order to simulate atmospheric turbulence in the lab we use phase plates. Phase plates are made from a transparent material, such as glass or plastic, and have a precise coating of clear acrylic spray paint on them; depending on how the coating of acrylic spray paint is applied affects how large or small $r_o$ will be \cite{Rampy}. In our case we want $r_o$ to be small, relative to our optical setup, in order to properly and accurately simulate atmospheric turbulence. Since the application of the acrylic paint effects $r_o$ we are able to produce new phase plates of various $r_o$’s until we have the technique to create proper atmospheric turbulence; an $r_o$ which follows a 5/3 power law. After the acrylic is applied to the phase plates we then characterize them and determine their $r_o$. This is generally accomplished, in simple terms, by first measuring the interference patterns, obtaining a surface, and measure the root mean square (RMS) variation of the surface. 

To acquire our data we use an interferometer, which after carrying out various steps which will be explained later, we can measure the phase and determine the $r_o$ of any phase plate of our choosing. An interferometer uses a single light source that is split into two beams and when polarized creates an interference pattern. By measuring an interference pattern we can recreate the surface of phase plates or a mirror and determined on a technique that measure the RMS variation in order to calculate a $r_o$. Using an interferometer allows us to improve the process of determining phases’ and $r_o$s’ with greater accuracy since we are able to do everything in a controlled environment and are able to repeat the process with various phase plates multiple times.

In order to confirm that our phase plates accurately simulate atmospheric turbulence we verify $r_o$ by making sure it fits Kolmogorov’s power law. Kolmogorov’s power law is a power law of 5/6, “a variance in temperature between two points a distance x apart” \cite{Beckers93}. When testing phase plates they may either not follow this power law or they may have an $r_o$ that is either far too high or far too low. Because of this we make new phase plates in order to be able to obtain the desired $r_o$ that accurately simulates atmospheric turbulence. 

\chapter{Methodology}
\section{Quadrature Polarization Interferometer (QPI) Intro}
\par
An interferometer is an instrument that uses one single light source split into two to create an interference pattern which can then be analyzed. The interferometer used in this experiment is based on the Mach-Zehnder Interferometer. The Mach-Zehnder Interferometer is an instrument used to measure and determine the phase shift between two collimated beams of light that are derived from one single light source \cite{M-Z}. The operation of a Mach-Zehnder interferometer goes as follows: the single light source is properly focused and collimated using a lens then using a beam splitter it is split into the two beams, the test and reference beams. From here the two separate beams are reflected using mirrors towards a second beam splitter where they are then registered by a single detector. The Mach-Zehnder interferometer allows for the light to travel through an equal optical length in both the test and reference beams which then results in constructive interference. Since the Mach-Zehnder interferometer is also a highly adaptable/configurable instrument, it is a great interferometer to derive our interferometer from.

Building off of a Mach-Zehnder interferometer, a phase shifting interferometry allows us to "measure the phase variation across the beam and then convert them into height variations across the sample" \cite{Wyant}. This phase difference is determined by measuring the intensity of the interference patterns from both cameras which we know to have a 90\textdegree{} phase change between them. This 90\textdegree{} phase change between them typically results from moving the mirror in the reference beam back and forth to change the path length which results in a phase change. A phase shifting interferometer uses the intensities across the interference frames, converts this to the phase variations, and converts this to the final step of height variations, optical path length, in order to recreate surface of the imaged object which in turn allows us to calculate the $r_o$ of said object.

For this thesis work we use a Quadrature Polarization Interferometer (QPI). A Quadrature Polarization Interferometer lets us measure the phase delay in phase plates which we can then use to determine the $r_o$ of the phase plate. QPI makes this possible because it is a type of phase shifting interferometer, like described above, which allows us to get two interference patterns simultaneously, which is phase shifted by 90\textdegree{}. The QPI setup used in this thesis is shown in the following figure:
\begin{figure}[ht]
  \centering
  \includegraphics[width=\textwidth]{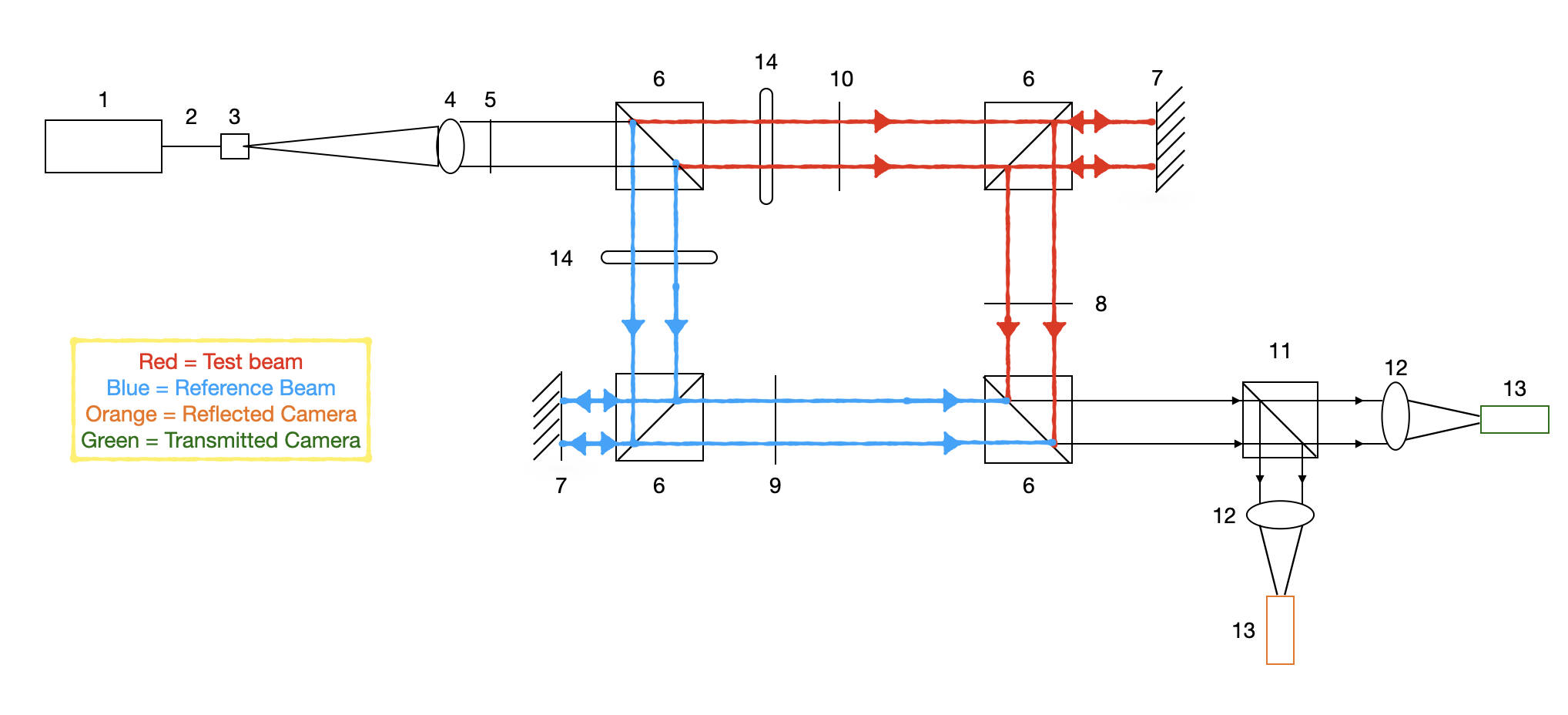}%
  \caption{Quadrature Polarization Interferometer Layout}
  \label{fig: QPI}
\end{figure}

\begin{table}[!htb]
\begin{center}
\scalebox{.6}{%
\begin{tabular}{||c|c||}\hline
\multicolumn{2}{||c||}{\textbf{QPI Components}} \\\hline\hline
  \textbf{Number} & \textbf{Component} \\[1ex]\hline
  1   &  HeNe Laser Source\\[1ex] \hline
  2   &  Optic Fiber Cable\\[01ex] \hline
  3   &  Cable to Laser Mount\\[01ex] \hline
  4   &  Thorlabs Lens f=400mm\\[01ex] \hline
  5   &  Linear Polarizer\\[01ex] \hline
 6   &  2" Non-Polarized Beam Splitter\\[01ex] \hline
  7   &  Mirror\\[01ex] \hline
  8   &  2" $\frac{\lambda}{2}$ Waveplate\\[01ex] \hline
  9   &  2" $\frac{\lambda}{4}$ Waveplate\\[01ex] \hline
  10   &  Phase Plate\\\hline
  11   &  11" Polarized Beam Splitter\\[01ex] \hline
  12   &  Thorlabs Lens f=100mm\\[01ex] \hline
  13   &  HT-5000-S Emergent Camera\\[01ex] \hline
  14   &  Motorized Flip Mount \\\hline
\end{tabular}}
\caption{Components Required for QPI}
\label{tab:QPI}
\end{center}
\end{table}
\pagebreak{}

 By using this method we can test more than just a phase plate; by removing the mirror in part 7 of Figure \ref{fig: QPI} and replacing it with a small deformable mirror we are able to test the deformable mirror after the phase plates have been characterized. Although I will not be testing a small deformable mirror, once QPI is complete and I have shown that it provides accurate/expected results for $r_o$ this is a possibility for the future.

As described in the Mach-Zehnder interferometer description above, a single light source is split into two beams, the test and reference beams. By introducing a quarter-wave plate into the reference beam or by changing the optical length of the reference beam, this allows for the measurement of phase difference or phase delay. This is the basis for how we will obtain the phase delays using QPI, by introducing a quarter-wave plate into the reference beam. QPI is a bit different as it introduces two more non-polarized beam splitters, a polarized beam splitter, two wave plates, and relocates both mirrors. The introduction of all the pieces allows us to better create constructive interference and to better detect phase delay.

\section{Assembly}
\subsection{Laser and Non-Polarizing Beam Splitters}

Using the same principles of the Mach-Zehnder interferometer but now adopted to QPI, our goal is to be able to characterize phase plates. In order to characterize phase plates we first need to derive the phase by using two polarization's that travel the same path length but it differring beams, a test beam and a reference beam. In QPI we do this with a HeNe laser (633 nm wavelength) however before introducing this HeNe laser we first need to do our alignments, starting with the laser alignment. The laser alignment is done with a handheld laser pointer; handheld laser pointers allow us to make better alignments as the beam is thin and focused which lets us make finer adjustments in our upcoming procedure. Using the handheld laser we first “mate” the laser to the work bench meaning we make sure the laser is parallel across the entire length of the work bench. Since the length of the work bench is far greater than the length the laser has to travel in the setup, this means that the laser is parallel and straight for the length of the QPI system.

“Mating” the laser to the bench is important to the setup, especially when we introduce the non-polarized beam splitters. The non-polarized beam splitters, as their name suggests, split the beam into two, with one going straight through and another splitting perpendicular from the first. The laser “mating” makes the introduction of the non-polarized beam splitters easier as we no longer have to worry about the laser alignment and can primarily focus on the placement of the beam splitters; placement meaning positioning, tip and tilt. Once all four of our non-polarized beam splitters have been properly placed, along with both mirrors which are placed in the same manner, we can then replace our handheld laser pointer with a HeNe laser in order to obtain a proper source rather than a point source.

We place our HeNe laser source at the same height at which we had our handheld laser pointer in such a manner that the center of the HeNe laser is also the center of the non-polarized beam splitters, as it was for the handheld laser. In order to obtain a proper “beam” of light we first need to place a lens into the path of our HeNe laser light source; the lens we chose has a focal length of 400 millimeters (mm). Placing the lens at ~400 mm away from the HeNe laser source, we then used a shear plate to determine where exactly the beam is properly collimated since this provides us with the exact/proper placement for the lens.

Once we had our main beam aligned and properly collimated we added a polarizer behind the focus lens that produced a linearly polarized beam which is parallel to the testbed/bench. Since we have already accurately aligned all four non-polarized beam splitters we can see how the beam travels through the setup as of this step. The linearly polarized beam is split into two at the first beam splitter creating what will be referred to as the test and reference beams. From there two other beam splitters, along with both mirrors, redirect the beams and allow for them to rejoin and merge together at the final and fourth beam splitter. After the two beams have been recombined, a smaller polarizing beam splitter is added to the system, following the same procedure as for the non-polarized beam splitters, so that the beam can split once more and separate into the two phases; these separated beams should be at an angle difference of 90\textdegree{}s providing us with the trig functions cos and sin. The polarized beam splitter should be aligned during the laser alignment process with the handheld laser pointer in order to obtain the best alignment results; if necessary it is possible to introduce it into the system and align it once the HeNe laser is in place although this is slightly harder to get the correct alignment but completely possible. I mentioned a phase difference between the two beams however as of this step in the procedure there is no phase difference. The phase difference between two beams is achieved by adding wave plates into the system.

\subsection{Wave Plates}

A retarder or wave plate is an optic which changes the polarization of the light that travels through it. In order to obtain the phase difference between the two beams we introduce two wave plates into the system, a half-wave plate and a quarter-wave plate into the test beam and reference beam respectively. By using a half-wave plate and a quarter-wave plate we are able to make one beam rotate by 45\textdegree{} and make the other beam circularly polarized respectively. When we input the polarizer at the start of QPI it polarizes the light so that it is parallel to the table/bench. The half-wave plate is placed in the test beam and takes this linearly polarized light and rotates it by 45\textdegree{} which gives us a polarized light that gives us sin and cos. However if we were to look at our cameras in this step we would see the same interference patterns on both cameras with no phase difference between the two, this is where the quarter-wave plate comes into play. Placing the quarter-wave plate in the reference beam, it takes the linearly  polarized light and converts it to circularly polarized light. Circularly polarized light means that as one trig function is at its maximum the other is at its minimum, meaning that there is a 90\textdegree{} phase difference between the trig functions, this is how we get the 90\textdegree{} phase shift between cos and sin that we see in the cameras. The final step lies in the polarized beam splitter which separates both trig functions by allowing only one polarized direction to transmit through it to one camera while the other polarized direction is reflected towards the other camera.

\begin{figure}[ht]
  \centering
  \includegraphics[width=\textwidth]{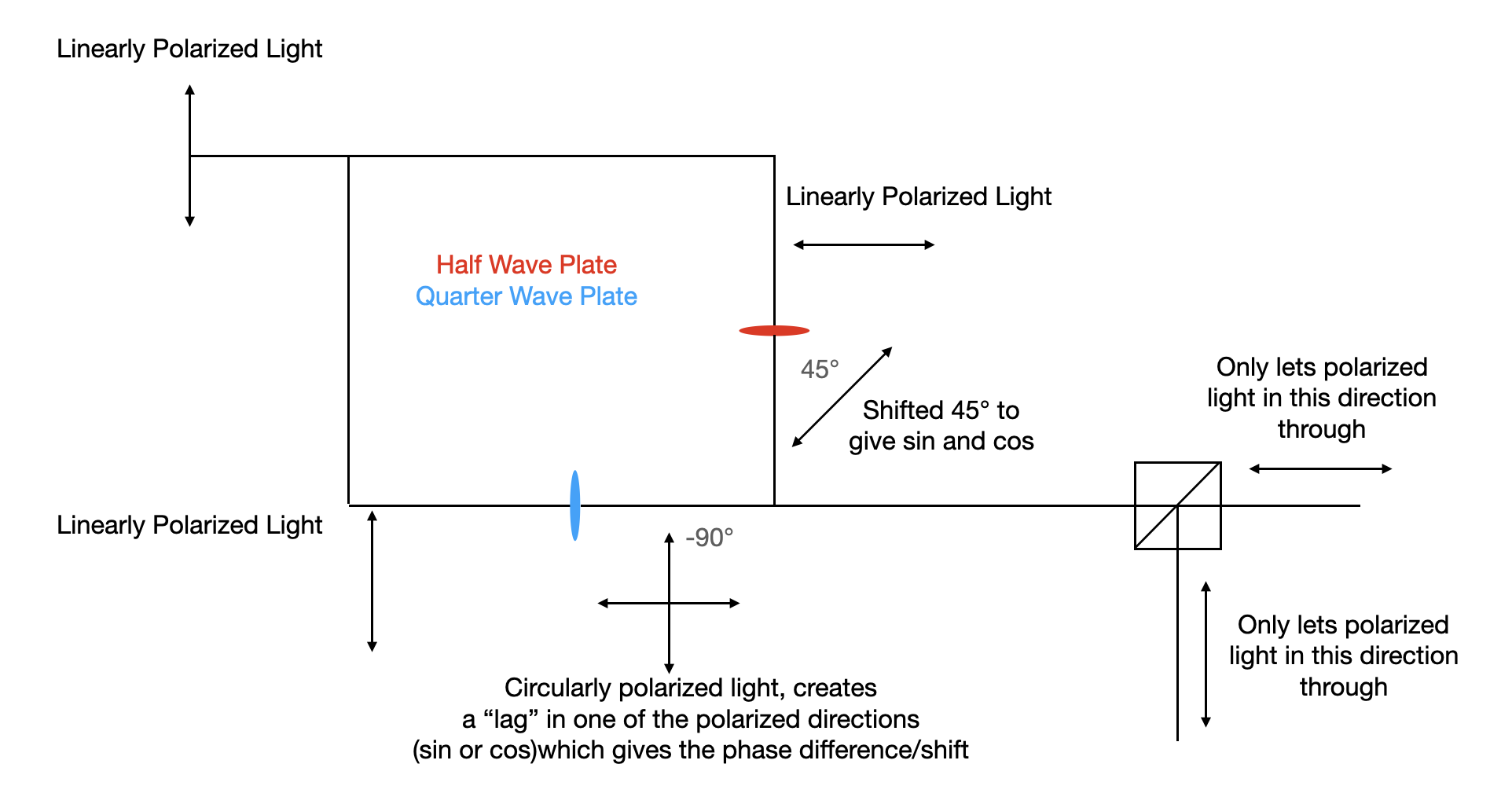}%
  \caption{Wave Plates Operation: The figure above demonstrates the manner in which the wave plates create the trig functions sin and cos and the phase difference between them.}
  \label{fig: WavePlates}
\end{figure}
\pagebreak{}
To make sure we set up the wave plates correctly we need to use a power meter as well as a second polarizer. A power meter allows us to determine the intensity of the beam, which allows us to set the wave plates accurately.

The wave plates are introduced into the system one at a time. We begin by inserting the half-wave plate, however before the half-wave plate is inserted into the system, we first want to place a second polarizer on the opposite side of the beam splitter; opposite with respect to where the half-wave plate will be placed. Once the second polarizer is set we then place the power meter directly behind it. Using the power meter, we first find the angle of maximum flux for the second polarizer. Once this angle is found we then insert the half-wave plate into the test beam path between in place number 8 of Figure \ref{fig: QPI}, by doing so the angle of maximum flux has now changed. In order for the half-wave plate to be aptly adjusted we want there to be a 45\textdegree{} angle change in the polarizers angle of maximum flux when the half-wave plate is inserted and absent; meaning if the angle of maximum flux of the polarizer is 25 degrees when the half-wave plate is missing, it should then be 70 degrees with the half-wave plate inserted. Since we know the angle of maximum flux for the polarizer when the half-wave plate is absent, we can do the following: by changing the angle of the polarizer by 45 degrees we can then focus only on finding the angle of maximum flux for the half-wave plate. Once this is complete, the half-wave plate has been adequately placed and we can move on to the quarter-wave plate.

The adjustment of the quarter-wave plate is done using the same equipment used for the half-wave plate, a power meter and a polarizer. We start by moving the power meter and the polarizer to opposite side of the non-polarized beam splitter from where the quarter-wave plate will be placed. We then rotate the polarizer to see the expected flux variation and confirm that the light is linearly polarized. By rotating the polarizer we are able to find two angles, the minimum flux and maximum flux angle. Once we have these two angles we insert the quarter-wave plate in the reference beam path in part number 9 of Figure \ref{fig: QPI}. We then want to find the angle for the quarter-wave plate that provides no change in flux for the minimum and maximum flux angles we had previously found for the polarizer. Meaning we want to find an angle for the quarter-wave plate that when we switch between the max and min angles on the polarizer there is no variation/change in their flux. After this is complete, the setup of the wave plates is complete and we remove the second polarizer as well as the power meter from the setup as they are no longer needed. 

\subsection{Focusing Lens’ and the Final Alignment}

Once the beam has traveled through the system and through the polarized beam splitter we need to focus the light to image it onto the cameras. In order to do this we will use lens’ that have a focal length of $\sim$100mm. Placing the lens’ in positions 12 of Figure \ref{fig: QPI} we adjust the height of them so that their center is at the same height as the center of the polarized beam splitter and place them at an equal distance from the polarized beam splitter. From here we begin constructing the setup for the cameras. In order to make adjustments easier, we mount the camera on 3 different movable bases. The 3 bases allow for us to move the camera small amounts in any direction, up and down, forwards and backwards, left and right. After mounting the bases to one another and finally the camera to them, we place them $\sim$100mm away from the lens’. 

From here we need to see the live feed of the cameras to be able to adjust their positioning properly. Following their setup instructions we connected them to the computer and were able to see their live feed. Next we place an iris in part number 10 of Figure \ref{fig: QPI}, where the phase plates will eventually go, and use a motorized mount to block the beam from traveling in the reference path. The iris aperture can be any size you wish but we chose a medium size aperture as it makes the next step slightly easier. Looking at the live feed of the cameras now we see that they are not focused or properly aligned; they have different magnifications, different positioning (imagining of light onto camera), and are unfocused. By using the iris as the reference we use the bases to move the camera until the iris aperture edge is seen as a sharp edge instead of a blurry edge on the live feed; this is when the focus is best. We do this for both cameras and once both cameras have the best focus we can manually adjust for, we switch over to adjusting for the magnification of the cameras. 

In order to adjust the magnification we need to change how close/far the lens is to the polarizing beam splitter; however by doing so it messes up the focus of the cameras. This means that as we correct the magnification we also need to adjust the positioning of the cameras in order to regain the focus. One of the best ways to do this is to try and place both lenses equally far away from the polarized beam splitter, which we did. From here we make slight adjustments until we are able to get very similar magnification and great focus on both cameras. After this process is complete, QPI’s setup has officially been completed and we can now begin testing.

\chapter{Data}

In this section we describe the processes of data acquisition with QPI. This includes calibrating the system, describing a data set, and a walk-through of how to take the images for said data set.

\section{Image Registration and Calibration}

In order to begin taking images, or data, we first need to register and check the pixel scale of each camera. To check calibration we make sure both cameras are centered the same and their images perfectly overlap with one another. This is accomplished by using a 1951 USAF resolution test chart:
\begin{figure}[ht]
  \centering
  \includegraphics[width=5cm]{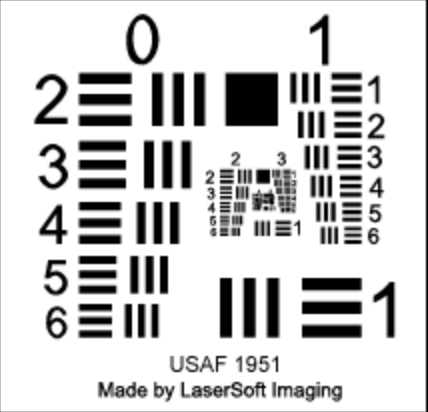}%
  \caption{1951 USAF Resolution Test Chart [Taken from Wikipedia "1951 USAF resolution test chart"]}
  \label{fig: USAF}
\end{figure}

By placing the test chart where the phase plate should go in part number 10 of Figure \ref{fig: QPI} and blocking the reference beam, to avoid creating interference, we are able to adjust the positioning of the lenses (up/down or left/right) until both test chart images are aligned over one another. This process is easier to complete if there is a program that displays the live feed of both cameras on top of each other so that we can make lens adjustments in real time.

\begin{figure}[h!]
  \centering
  \subfloat{{\includegraphics[width=5cm]{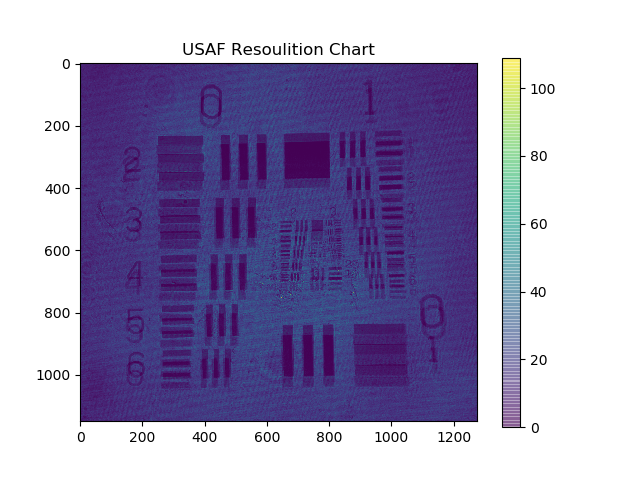} }}%
  \qquad
  \subfloat{{\includegraphics[width=5cm]{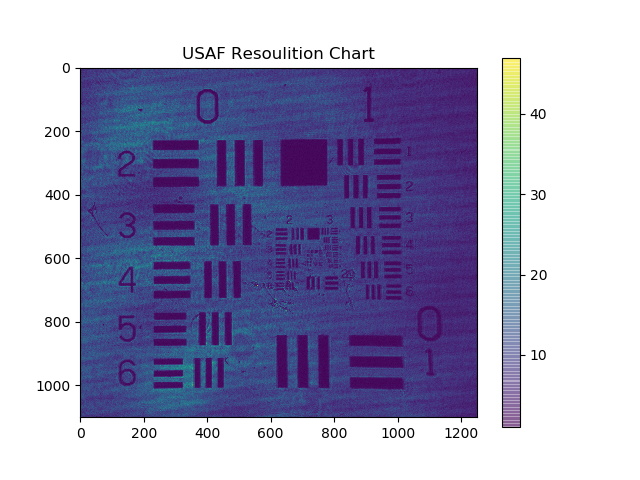}}}%
  \label{fig:USAF Comparison}%
  \caption{USAF Resolution Chart Comparison. These images show the two camera images added together. Left image the test charts are not aligned therefore the cameras have to be adjusted; the right images shows the test charts correctly aligned.}
\end{figure}
\pagebreak{}

Once the resolution test chart shows us that the images overlap over one another well enough we move on to our next and final calibration step. We take an image of the non-plate interference pattern, using either camera, and plot the x-axis of the image. By doing this we are able to see which, if any, of the pixels are saturated. Saturation occurs when too much light is being registered by a pixel and if any pixels are saturated, or if pixels are close to being saturated, we lower the intensity of the HeNe laser. 
\begin{figure}[ht]
  \centering
  \subfloat{{\includegraphics[width=5cm]{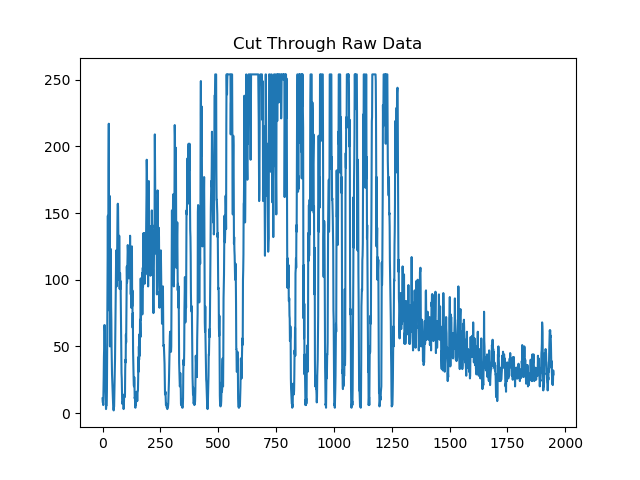} }}%
  \qquad
  \subfloat{{\includegraphics[width=5cm]{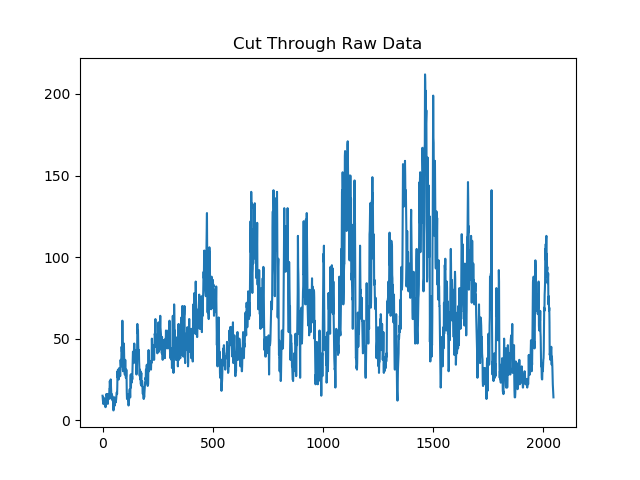}}}%
  \label{fig:Cut Raw}%
  \caption{Saturation Cuts. Left images has peaks which are cut off because the pixels are heavily saturated whereas the right images the beam intensity has been adjusted and the pixels are no longer saturated.}
\end{figure}

From here we need to determine pixel size to mm and we do this by looking at the cameras we are using in QPI, which are HT-5000-SM Emergent cameras. We take an image of the resolution test chart, with either camera as long as the reference beam is blocked, and analyze the saved image. By looking up the standard pattern and matching our selected group and element we are able to look that up on a table which displays the width of 1 line:

Using these references and using group 0 element 2 we see that the width of one of those lines is 445.45 microns; using our image we determine that the width of that same line is $\sim$30 pixels. With these numbers we calculate that 1 pixel is ~14.8 microns; this is the conversion that will be used when calculating $r_o$.

\begin{figure}[ht]
  \centering
  \includegraphics[width=\textwidth]{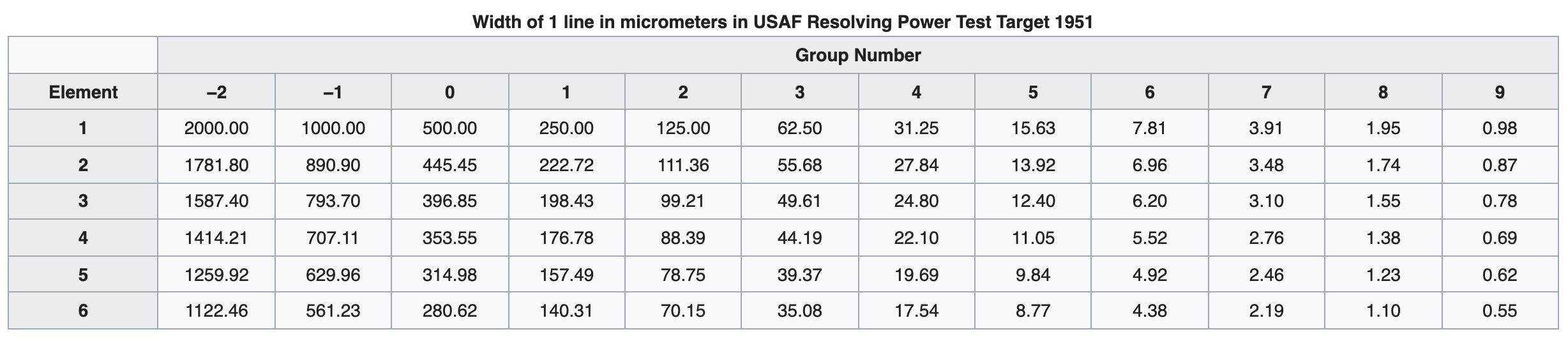}%
  \caption{Width of 1 line for USAF Chart [Taken from Wikipedia "1951 USAF resolution test chart"]}
  \label{fig: USAF Width}
\end{figure}

\section{Data Acquisition Process}
\subsection{Defining a Data Set}

With the calibration and pixel conversion now done we can proceed to take the images needed to determine the $r_o$, this set of images we refer to as a data set. A data set consists of 6 total images, 3 image sets; an image set being two images for: interference images, test beam images (block the reference beam), and reference beam images (block the test beam). These images are the actual measurements we will take where the interference images are the ones we want and the test beam and reference beam images are needed in order to scale the flux properly. In total this makes 6 images, 3 images taken with the transmitted camera and 3 images taken with the reflected camera; this is a data set.

When going to take a data set it is important to have both cameras synced with one another so that when an image is saved both cameras capture the image simultaneously, this is most important when capturing the interference images. If the cameras are not synced and do not capture the images simultaneously then because of vibration, the interference lines on the images will not be 100\% aligned which may cause a small issue later on when analyzing the data set. Camera synchronization ensures that we obtain very accurately aligned images and rules out the effects of vibration and the slight fluctuations in beam intensity. 

\subsection{Taking the Images}

In order to begin taking a data set we first need to make sure that the alignment is the same and nothing has been moved, either purposefully or accidentally. If QPI has been slightly adjusted or accidentally moved, we use the USAF resolution test chart in order to verify the camera images are aligned correctly over one another. If using the Graphical User Interface (GUI)  software that the cameras came with it is also good practice to verify that the pixel format for the cameras are in mono 12 instead of mono 8; if using a GUI that was built for easier use like ours, this is something that should be predefined but can be checked and modified if necessary.

After verifying the camera alignment we then proceed to take the data set images. We place the phase plate that we wish to measure in part number 10 of Figure \ref{fig: QPI} and make sure that both the test beam and reference beam blockers are both up; up meaning the path(s) are not being blocked and down meaning the path(s) are being blocked. Using the cameras’ GUI which allows us to save the image from both cameras simultaneously we save the first image set, the interference pattern images. Now by moving the test beam blocker down we save and have our next set of images, the reference beam images (test beam blocked images). Finally we move the test beam blocker up and the reference beam blocker down and proceed with saving this set of images, the test beam images (reference beam blocked images). After this set of images is taken we now have our full data set for the phase plate such as in the following figure:

\begin{figure}[ht]
  \centering
  \subfloat{{\includegraphics[width=5cm]{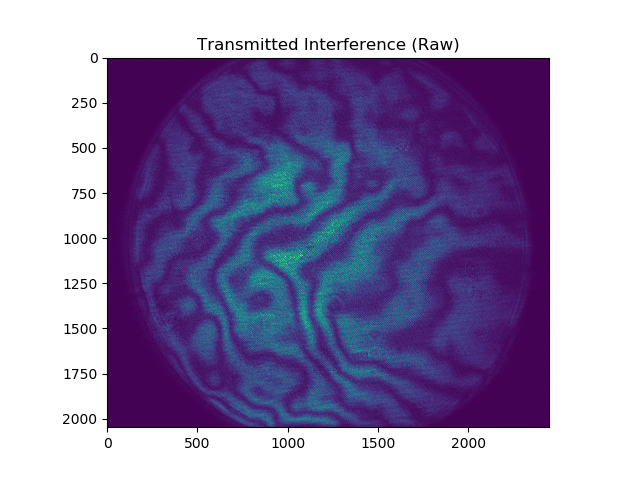} }}%
  \qquad
  \subfloat{{\includegraphics[width=5cm]{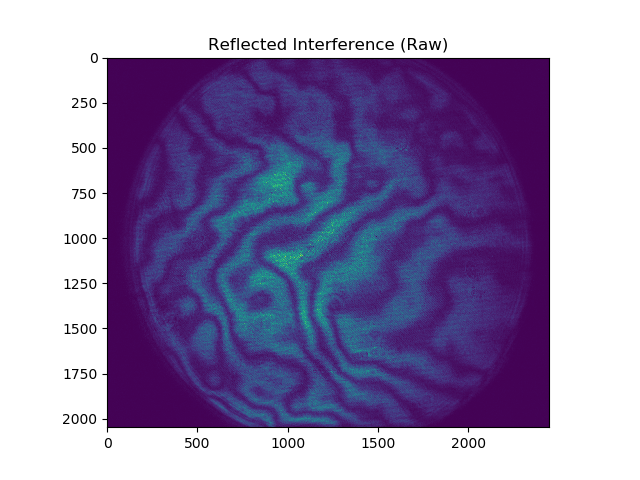}}}%
\end{figure}
\begin{figure}[ht]
  \centering
  \subfloat{{\includegraphics[width=5cm]{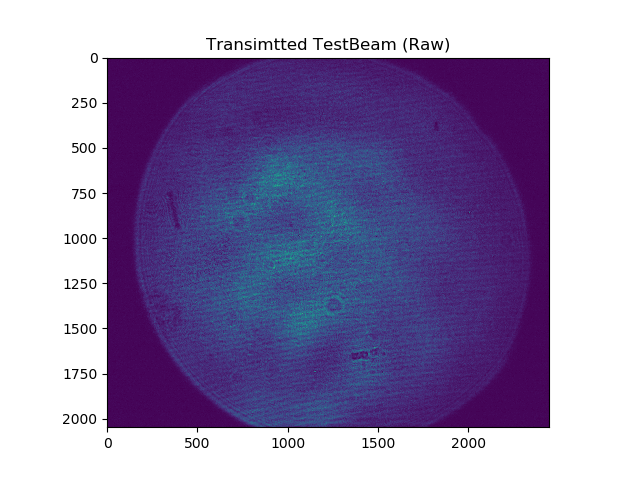} }}%
  \qquad
  \subfloat{{\includegraphics[width=5cm]{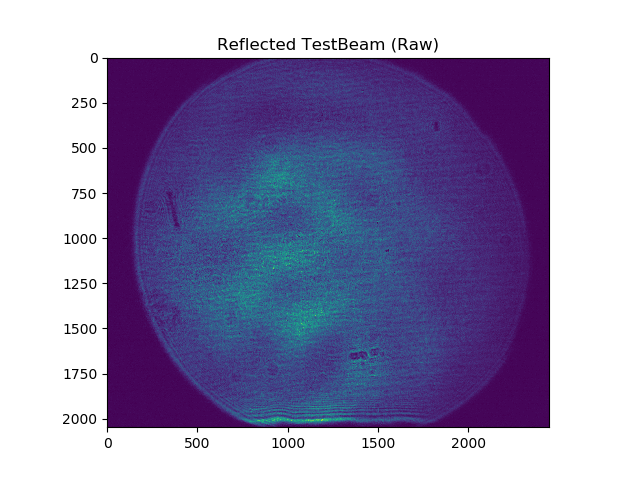}}}%
\end{figure}
\begin{figure}[ht]
  \centering
  \subfloat{{\includegraphics[width=5cm]{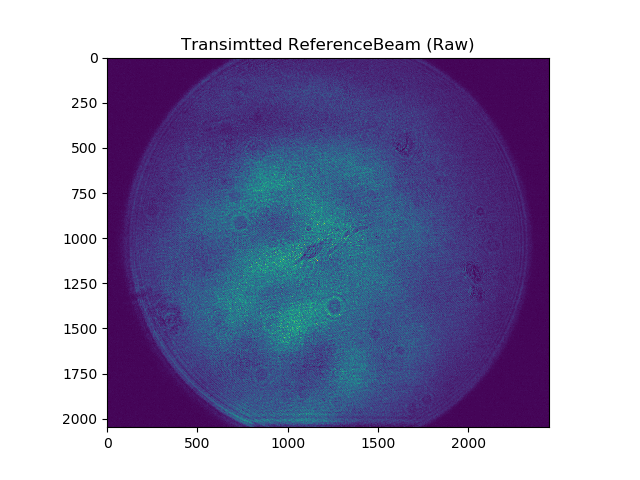} }}%
  \qquad
  \subfloat{{\includegraphics[width=5cm]{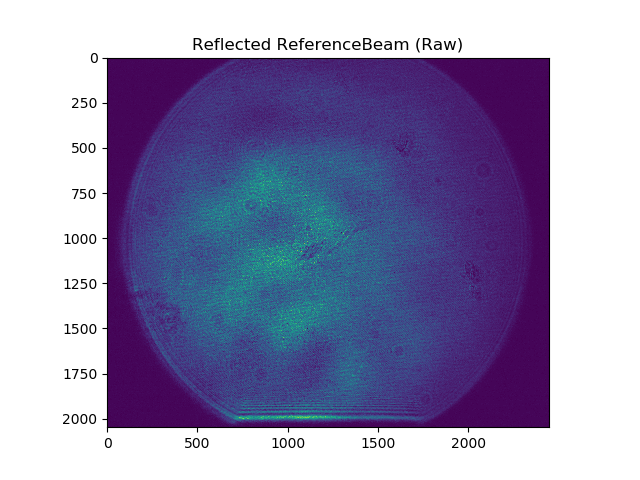}}}%
  \label{fig: Data Set}
  \caption{Complete Data Set. A data set consists of 6 total images; 3 image sets. One row in the figure above is considered an image set. All images are required in order to be able to determine $r_o$.}
\end{figure}
\pagebreak{}

\chapter{Results}
\section{Image Analysis}

In this section we walk-through the process of analyzing the images. This section includes describing how we start from the acquired raw data images to deriving the optical path delay across a phase plate, and our final method for calculating $r_o$ for said phase plate. 

All the image analysis as well as the $r_o$ calculations explained in the following sections were accomplished using Python 3.0.

\subsection{Image Processing}

Once we have a full data set for a phase plate we can then proceed to analyze it and determine its $r_o$. Since we have already made sure that the images are aligned using our own custom GUI and the 1951 USAF resolution test chart we have already confirmed that the two cameras images are aligned. Because of this there is no image correction that we have to apply through the coding process which means that we can begin the analysis. 

In order to be able to determine $r_o$ there are a number of processes we must first complete. The process begins by taking the raw data set images and normalizing them in order to calculate the maximum flux for each pixel, a value that should range between 0-2. Normalizing usually results in values between 0-1 however we expect a range between 0-2 because 1 here defines the approximate light expected from one beam path and since we have 2 beam paths we therefore expect a range from 0-2. We normalize the image by using the equation: 
\begin{equation}
    I_{Norm} = \frac{I_{Int}}{(I_{ref} + I_{Test})}
\end{equation}
where $I_{Norm}$ is the normalized image, $I_{Int}$ in the interference image, $I_{Ref}$ is the reference image, and $I_{Test}$ is the test image.
\begin{figure}[ht]
  \centering
  \includegraphics[width=\textwidth]{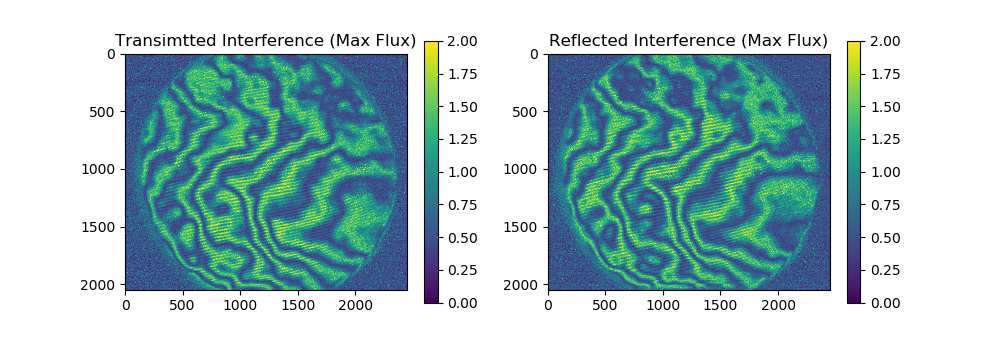}%
  \caption{Max Flux for Both Cameras. Normalization of the raw images for transmitted and reflected cameras.}
  \label{fig: Max Flux}
\end{figure}
\pagebreak{}

Once we have the maximum flux we need to convert this to intensity. Determining the intensity requires the modification of the normalization through the following equation: 
\begin{equation}
    I_{Sin} = I_{Norm} - 1
\end{equation}
which should provide an intensity that ranges between -1 and 1. By doing so we are converting to the expected ranges of the tragicomic functions sin and cos.
\begin{figure}[ht]
  \centering
  \includegraphics[width=\textwidth]{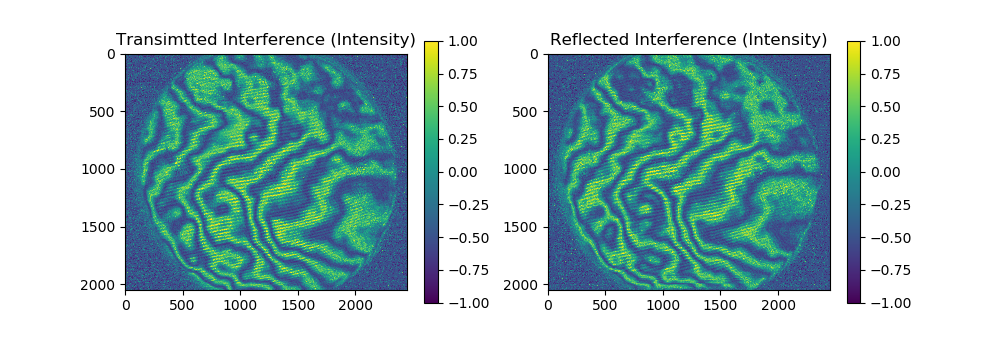}%
  \caption{Max Intensity for Both Cameras. We have to adjust the normalization in order to obtain values that will work for the trig functions Sine and Cosine.}
  \label{fig: Intensity}
\end{figure}

It is important to note that there are two cameras meaning that both these equations will have to be calculated twice per data set; once for the transmitted camera and once for the reflected camera. 

For example if the transmitted images are used when defining $I_{Sin}$ then the reflected images should be put through the exact same equations except should be named $I_{Cos}$ instead of $I_{Sin}$ because of a phase shift. This phase shift between the two is ideally 90\textdegree, thus their names $I_{Sin}$ compared to $I_{Cos}$. This ideal, perfect 90\textdegree{} phase shift is only possible if the the images in an image set where taken simultaneously, if there was no vibration in the optics, and if there was no fluctuation in beam intensity; if these variables cause the phase shift to vary form 90\textdegree{} significantly then this will lead to errors in the next step. Because of this it is very important to align QPI to the best of our ability and try to make the cameras take their images simultaneously. 

Once we have the intensities, $I_{Cos}$ and $I_{Sin}$, for both cameras we then proceed to calculate the phase values. Plugging in both intensities into the equation:
\begin{equation}
    Phase = 2 * \arctan{(\frac{I_{Sin}}{I_{Cos}})}
\end{equation}
we are able to calculate the phase values which should range between -pi and pi. 
\begin{figure}[ht]
  \centering
  \includegraphics[width=9cm]{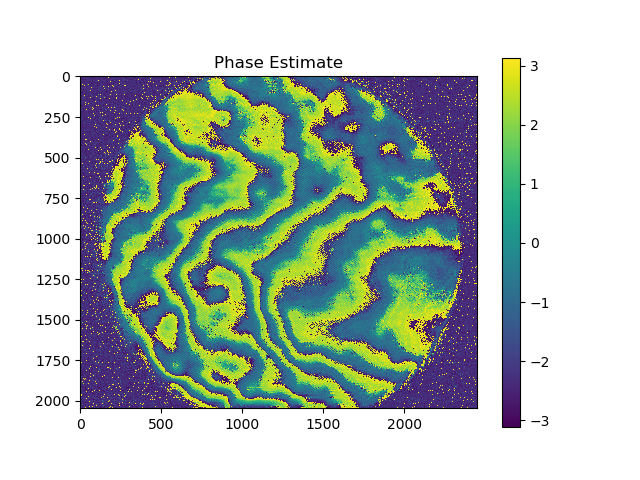}%
  \caption{Phase Estimate for Phase Plate}
  \label{fig: Large Phase}
\end{figure}

If we had the ideal 90\textdegree{} phase shift between both cameras then if we were to look at a cut of the plot we would have an asymmetrical sawtooth shaped graph. The phase shift between the cameras is not a perfect 90\textdegree{} which may be a result of vibrations in the optics and/or because we do not have perfect sync between our two cameras; they, at the moment, cannot take images simultaneously. Although the frame sync difference is already small in order to obtain better synchronization, and hopefully the ideal 90\textdegree{} phase shift, we want the frame sync difference to be .5-1ms. 

After obtaining the phase values we take those values and unwrap them. Unwrapping releases the constraint that the phase values had before, from - pi to pi, and allows for the resulting values to be larger than a difference of 2*pi. Phase unwrapping takes the phase values and unwraps it into a surface; this unwrapped surface should be the surface of what we imaged, in this case the surface of the phase plate. The unwrapping algorithm works by looking for discontinuities and either adding or subtracting 2pi from all values that follow in order to correct for this discontinuity.
In simpler terms the phase unwrapping is the variation of path length from one area to another hence any light that travels through more paint [of the phase plate] than air will travel a longer distance which we will see in the unwrapped image. This is what the phase unwrapping algorithm determines and based on this it recreates the surface of the object which the light passed through, in this case the phase plate. After attempting various python methods of unwrapping we settled on:
\begin{equation}
    Unwr\_Phase = unwrap\_phase ( Phase)
\end{equation}
where Unwrap\_Phase is a module imported from a python package named skimage \cite{HerraezUnwrap}. Although this unwrapping algorithm is not perfect, it works remarkably well and provides far less errors than any other unwrapping method we used. 
\begin{figure}[ht]
  \centering
  \includegraphics[width=9cm]{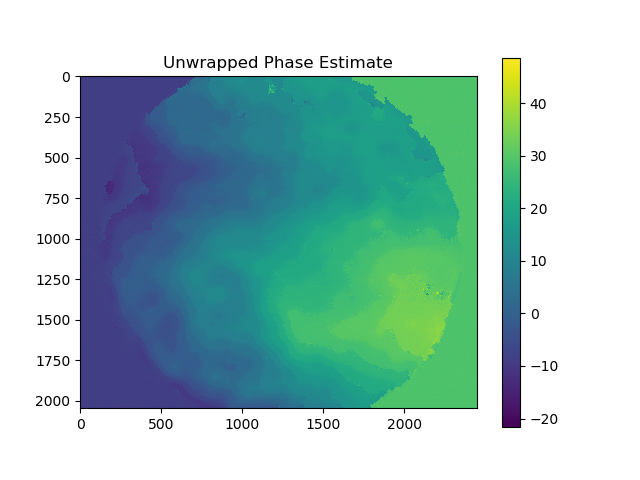}%
  \caption{Unwrapped Phase Estimate for Phase Plate}
  \label{fig: Large Phase Unwrap}
\end{figure}

Once we unwrap the phase and have recreated the surface of the phase plate we have finished the image processing process and can proceed with determining the $r_o$.

\subsection{Calculating $r_o$}

Calculating an accurate $r_o$ via the unwrapped phase is another procedure in itself. The basis for how we calculate the $r_o$ for these phase plates was shown by \cite{Beckers93}:
\begin{equation}
    \phi_{rms} = 0.162(D/r_o)^{5/6} 
\qquad{[waves]}
\end{equation}
where $\phi_{rms}$ is the root mean square for phase variation, D is the diameter of the aperture, and$r_o$ is atmospheric turbulence. However the equation above has the rms phase value in units of waves whereas we have it in units of radians thus we multiply the equation above by 2pi radians and get:
\begin{equation}
    \phi_{rms} = 0.162(D/r_o)^{5/6} (2\pi) 
\qquad{[radians]}
\end{equation}
From here we take the equation above and rearranging to solve for $r_o$ we obtain:
\begin{equation}
    r_o = \frac{D}{(\phi_{rms}/(0.162 * 2\pi))^{6/5}} 
\qquad{[cm]}
\label{eq: final}
\end{equation}

Since our $\phi_{rms}$ values are in radians and so is 2 pi, the radians cancel leaving us with units of length which is expected. An important note about the equation above is that it assumes that all the phase plates we test/measure follow Kolmogorov’s 5/6 power law. Once we have our equation we can begin calculating for $r_o$. In order to calculate $r_o$ we need to determine both the unknowns in the equation above. As mentioned previously $\phi_{rms}$ is the root mean square for phase variation meaning that it is the RMS for an aperture of [unwrapped] phase values; the diameter of this aperture is D.

The process for calculating $\phi_{rms}$ goes as follows: choose a pixel and select all the pixels around it within an aperture of diameter D. The process continues by taking all those pixels and taking the standard deviation of all their unwrapped phase values, this is $\phi_{rms}$, however this is not the correct $\phi_{rms}$ yet. Our HeNe laser is at a wavelength of 633 nm however $r_o$ is generally calculated at a wavelength of 550 nm meaning we must take our $\phi_{rms}$ and multiply it by (633/550). From here we now have the correct $\phi_{rms}$ value.

Once we have $\phi_{rms}$ we take it and the diameter size D we used to calculate it and input both of these values into equation \ref{eq: final}. In doing so we obtain the $r_o$ for that specific spot, for that aperture size. Now that we can calculate the $r_o$ for a specific aperture we need to decide how we will calculate $r_o$ for the entirety of the plate. In order to get a good sense of what $r_o$ is across the entirety of the plate we decided to to select multiple random pixels across the phase plate, choose a small aperture size, calculate their individual $r_o$’s and average them all in order to get one final averaged $r_o$ which will represent the $r_o$ of the entire phase plate. We chose this method because by averaging numerous $r_o$’s of all the same aperture but different locations on the phase plate we calculate a more refined and averaged $r_o$ than if we were to calculate one $r_o$ that has an aperture size of the entire, if not most, of the phase plate. 
\begin{figure}[ht]
  \centering
  \subfloat{{\includegraphics[width=5cm]{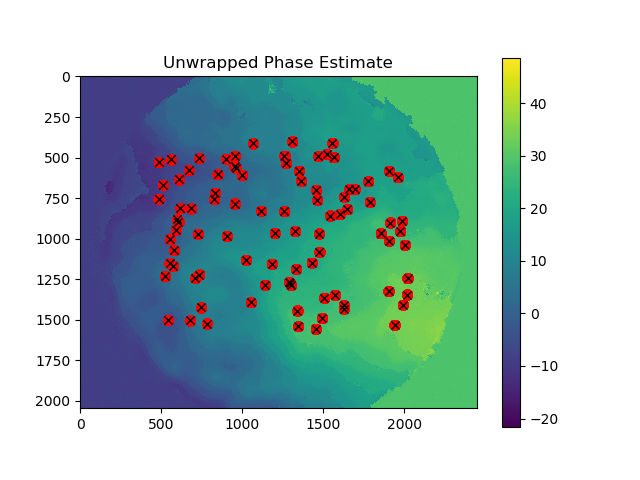} }}%
  \qquad
  \subfloat{{\includegraphics[width=5cm]{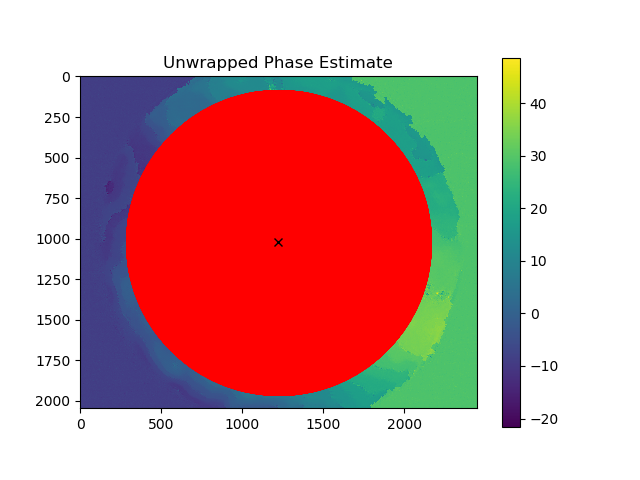}}}%
  \caption{Comparison of $r_o$ Calculations. The left method calculates $r_o$ for multiple random pixels and averages them out in order to obtain a good averaged $r_o$ for the phase plate which was $r_o$=1250microns. The right image calculates $r_o$ once by measuring the entire phase plate however the final $r_o$ tends to be inaccurate, as for the right image it gives $r_o$ = 1600microns}
  \label{fig: Multi and Single}
\end{figure}
\pagebreak{}

We have explained how to calculate the $r_0$ of any phase plate, all images above were examples of large $r_o$ and here is an example of a smaller $r_o$:
\begin{figure}[ht]
  \centering
  \includegraphics[width=\textwidth]{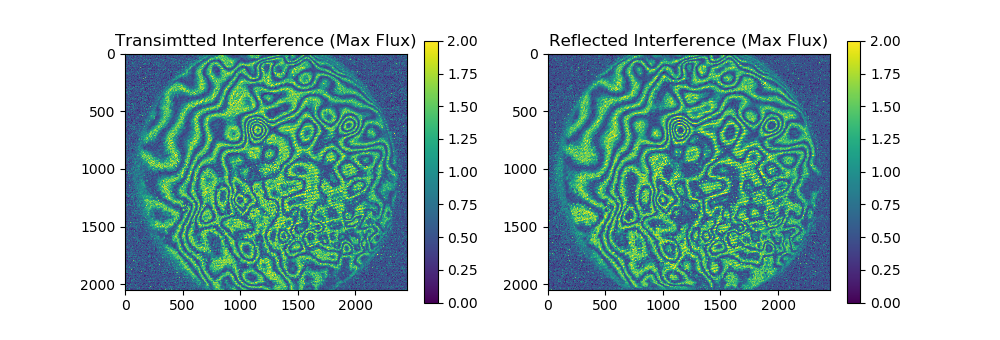}%
\end{figure}
\begin{figure}[ht]
  \centering
  \includegraphics[width=\textwidth]{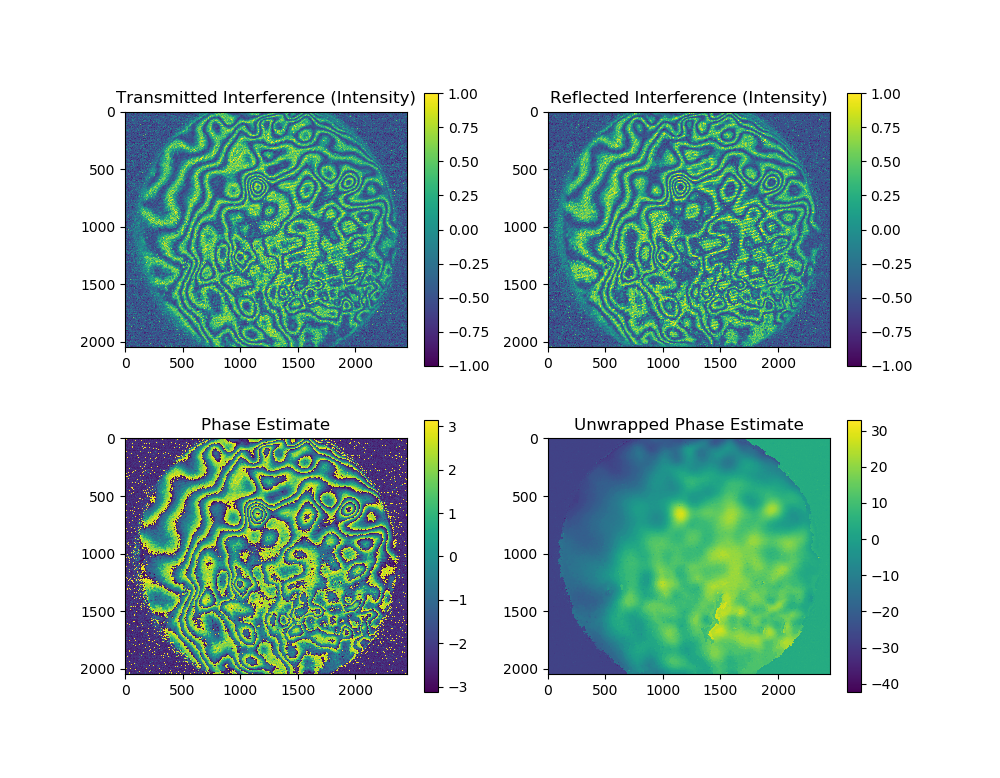}%
  \caption{Data Set for Smaller $r_o$}
\end{figure}
\begin{figure}[h!]
  \centering
  \subfloat{{\includegraphics[width=5cm]{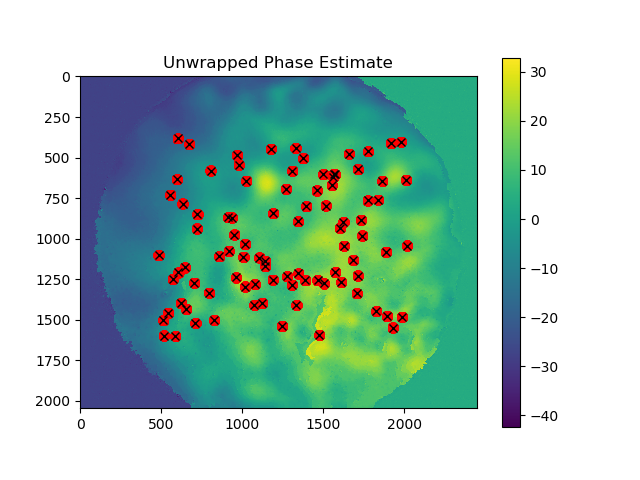} }}%
  \qquad
  \subfloat{{\includegraphics[width=5cm]{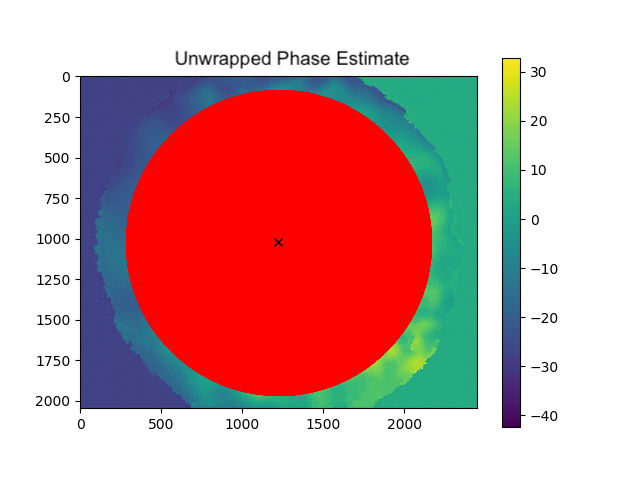}}}%
  \caption{Comparison of $r_o$ Calculations: The method shown in the left image calculates $r_o$ to be 650 microns while the method in the right image calculates $r_o$ to be 1380 microns.}
\end{figure}
\pagebreak{}

\chapter{Conclusion}
\section{Summary}

This thesis work required setting up a Quadrature Polarization Interferometer which is located in the Lab of Adaptive Optics at the University of Santa Cruz. This thesis was the product of wanting to analyze variations in phase delays across phase plates that simulate atmospheric turbulence in order to demonstrate that we can characterize phase plates accurately and consistently. The characterization of phase plates was accomplished by using QPI to examine the interference patterns between two separate paths and measure their pathlength variations. We were able to refine and create an algorithm that determines Fried’s parameter, $r_o$, of any phase plate we wish to measure using QPI and python code we developed. We did this by taking the raw images and converting them to determine their maximum flux, then their intensity, then their phase value, and finally their unwrapped phase values. The unwrapped phase values recreated the surface of the phase plate and we were able to calculate $r_o$ by analyzing the path length variations using the unwrapped phase values.

\section{Future Direction}

Now that QPI is set up, the LAO can now use it to measure phase plates or even test a small deformable mirror as I mentioned before. Future work will most likely entail attempting to get an even better alignment for QPI however the main focus will be the creation of phase plates. A side goal during this thesis was creating a process that provides a consistent $r_o$, i.e. if we want an $r_o$ of .5mm we want a phase plate creation process that consistently results in an $r_o$ of .5mm. Since we have now developed an algorithm that quickly calculates $r_o$ for any phase plates, we now create and test phase plates quickly, efficiently, and effectively.

\bibliography{Refs}   

\begin{thebibliography}{1}

\bibitem{Rampy}
{Rampy}, R., {Gavel}, D., {Dillon}, D., and {Thomas}, S., ``{Production of
  phase screens for simulation of atmospheric turbulence},'' {\em ao}~{\bf 51},
   8769 (Dec. 2012).

\bibitem{Beckers93}
{Beckers}, J.~M., ``{Adaptive Optics for Astronomy: Principles, Performance,
  and Applications},'' {\em araa}~{\bf 31},  13--62 (Jan. 1993).

\bibitem{M-Z}
{Zetie}, K.~P., {Adams}, S.~F., and {Tocknell}, R.~M., ``{TEACHING PHYSICS: How
  does a Mach-Zehnder interferometer work?},'' {\em Physics Education}~{\bf
  35},  46--48 (Jan. 2000).

\bibitem{Wyant}
{Wyant}, J.~C., ``{Dynamic Interferometry},'' {\em Optics \& Photonics
  News}~{\bf 14},  36 (Apr. 2003).

\bibitem{HerraezUnwrap}
{Herraez}, M.~A., {Burton}, D.~R., {Lalor}, M.~J., and {Gdeisat}, M.~A.,
  ``{Fast two-dimensional phase-unwrapping algorithm based on sorting by
  reliability following a noncontinuous path},'' {\em ao}~{\bf 41},  7437--7444
  (Dec. 2002).

\end{thebibliography}
\bibliographystyle{spiebib}   

\end{document}